\documentclass[11pt]{article}
\usepackage{comment}
\usepackage[hidelinks]{hyperref}
\usepackage{enumitem}
\usepackage{amsmath,amssymb,epsf,cite,graphicx,subfigure,physics}
\usepackage{amsmath,amsthm, tikzsymbols, dsfont, amsfonts}
\usepackage{algorithm}
\usepackage{algorithmic}
\usepackage{forloop}
\usepackage{comment}
\usepackage{xcolor}
\usepackage{graphicx}
\usepackage{dsfont}
\usepackage{soul}

\allowdisplaybreaks

\numberwithin{equation}{section}

\definecolor{airforceblue}{rgb}{0.36, 0.54, 0.66}
\newcommand{\beq}{\begin{equation}}
\newcommand{\eeq}{\end{equation}}

  \theoremstyle{definition}

\usepackage{graphicx}
\usepackage{epsfig}
\usepackage{amsfonts}
\usepackage{amscd}
\usepackage{latexsym}
\usepackage{amsmath,amssymb}
\usepackage{verbatim}
\usepackage{setspace}
\usepackage{color}
\usepackage{fancyhdr}
\usepackage{cite}
\usepackage{hyperref}
\usepackage{tensor}
\usepackage{xfrac}
\usepackage{comment}

\usepackage[textheight=9in, textwidth=6.5in, letterpaper]{geometry}

\numberwithin{equation}{section}

\def\0{{(0)}}
\def\1{{(1)}}
\def\2{{(2)}}

\def\co{{\cal O}}

\def\<{\langle }
\def\>{\rangle }

\newcommand{\av}[1]{\left\langle#1\right\rangle}

\newcommand{\pd}{\,\partial}

\newcommand{\bea}{\begin{eqnarray}}
\newcommand{\eea}{\end{eqnarray}}
\newcommand{\be}{\begin{equation}}
\newcommand{\ee}{\end{equation}}
\newcommand{\ba}{\begin{aligned}}
\newcommand{\ea}{\end{aligned}}

\def\be{\begin{equation}}
\def\ee{\end{equation}}
\def\beq{\be\begin{array}{c}}
\def\eeq{\end{array}\ee}

\def\4{_{\rm 4}}
\def\3{_{\rm 3}}

   \makeatletter
  \let\over=\@@over \let\overwithdelims=\@@overwithdelims
  \let\atop=\@@atop \let\atopwithdelims=\@@atopwithdelims
  \let\above=\@@above \let\abovewithdelims=\@@abovewithdelims
\renewcommand\section{\@startsection {section}{1}{\z@}%
                                   {-3.5ex \@plus -1ex \@minus -.2ex}
                                   {2.3ex \@plus.2ex}%
                                   {\normalfont\large\bfseries}}

\renewcommand\subsection{\@startsection{subsection}{2}{\z@}%
                                     {-3.25ex\@plus -1ex \@minus -.2ex}%
                                     {1.5ex \@plus .2ex}%
                                     {\normalfont\bfseries}}

\usepackage{tikz}
\usepackage{slashed}
\usetikzlibrary{calc}   
\usetikzlibrary{topaths}
\usetikzlibrary{decorations}
\usetikzlibrary{decorations.pathmorphing}

\begin{document}
\baselineskip=15.5pt
\pagestyle{plain}
\setcounter{page}{1}

\begin{center}
$$$$
{\LARGE \bf Cosmic ER \!=\! EPR in dS/CFT}
\vskip 1cm

\textbf{Jordan Cotler and Andrew Strominger}

\vspace{0.5cm}

{\it Harvard Society of Fellows, Cambridge, MA 02138 USA \\}
{\it Harvard University, Cambridge, MA 02138 USA \\}

\vspace{0.3cm}

{\tt jcotler@fas.harvard.edu, andrew$\_$strominger@harvard.edu}

\medskip

\end{center}

\vskip1cm

\begin{center}
{\bf Abstract}
\end{center}
\hspace{.3cm} 
In the dS/CFT correspondence, bulk states on global spacelike slices of de Sitter space are dual to (in general) entangled states in the tensor product of the dual CFT Hilbert space with itself.  We show, using a quasinormal mode basis, that the Euclidean vacuum (for free scalars in a certain mass range) is a thermofield double state in the dual CFT description, and that the global de Sitter geometry emerges from quantum entanglement between two copies of the CFT. Tracing over one copy of the CFT produces a mixed thermal state describing a single static causal diamond.

\newpage

\tableofcontents
\section{Introduction}

The so-far-imprecisely-formulated holographic principle posits that a bulk quantum theory of gravity has a dual representation as a quantum theory without gravity living on the boundary of  spacetime. In some cases, the emergence of the bulk spacetime geometry can be seen from the structure of quantum entanglement \cite{Ryu:2006bv,VanRaamsdonk:2010pw,Maldacena:2013xja,Almheiri:2014lwa}. A particularly clean example \cite{Garfinkle:1993xk} is the case of  two widely separated, pair-created black holes in a pure state. The  quantum correlations between the two black hole microstates cause large correlations between scalar fields at widely separated opposing points near the pair of black hole horizons,  resembling the correlations between widely separated Einstein-Podolsky-Rosen (EPR) photons. These large correlations  have a geometric interpretation as an Einstein-Rosen (ER) bridge connecting the two black holes. This situation was aptly summarized by the slogan `ER\,=\,EPR' \cite{Maldacena:2013xja}. 
 
The most completely understood example is in the stringy context of AdS/CFT, in which bulk AdS gravity is dual (when there are no horizons) to a CFT living on the AdS boundary at spatial infinity and bulk gravity states are directly identified with the boundary CFT states.  However, it was shown~\cite{Maldacena:2001kr} that the AdS black hole geometry, which has two boundaries, is an entangled  pure thermofield double state in the tensor product of the two CFT Hilbert spaces. In this example  the ER bridge connecting the two asymptotic exterior regions, as well as (more mysteriously) the black hole interior, can be understood as regions of spacetime which emerge from the quantum entanglement of the two boundary CFTs. 
 
In this paper we push this line of thought  one step further and argue that an entire de Sitter universe can emerge from quantum entanglement between two boundary CFTs. This is realized in the context of dS/CFT\cite{Strominger:2001pn}, where bulk states are dual to entangled sums of tensor product states in two boundary CFTs \cite{Ng:2012xp}. Quantum states in the CFT$_{d-1}$ dual to dS$_d$ live on $S^{d-2}$s in the $S^{d-1}$ at ${\cal I}^+$ (or ${\cal I}^-$). Each such $S^{d-2}$ is the boundary of a `northern' and `southern' pair of families of spacelike hypersurfaces whose union is a global spacelike $S^{d-1}$ slice in dS$_d$. A CFT$_{d-1}$ state is thus associated to each $S^{d-2}$ and a global bulk state is an entangled sum over pairs of such states. 
 
This setup is tailor-made for the emergence of a bulk spacetime from quantum entanglement between a pair of boundary CFTs.  We show for a free scalar in a certain mass range that the Euclidean vacuum for global dS can only be expressed in the boundary representation as an entangled, pure thermofield double state. We further show that one copy of the CFT can describe a (say) northern static patch of de Sitter space, most simply in a mixed thermal state. This thermal state can be purified with a second `southern' copy of the Hilbert space. The required entanglement produces the ER bridge between the two patches as well as the past and future Milne regions -- completing the bulk dual to a global de Sitter spacetime.
 
A key ingredient in our construction is the use of a quasinormal mode (QNM) basis for the scalar field, which was shown in~\cite{Jafferis:2013qia} to be complete for dS (for a suitable choice of function space).  These modes are very natural from the point of view of the dual CFT, as they organize into highest (and lowest) weight states and their descendants (ascendants). However, the QNMs unconventionally have imaginary frequency, rendering unitarity obscure and introducing other peculiar but interesting features. One important such feature which we derive in Sections~\ref{Sec:S4} and~\ref{Sec:S5} is that the reduced density matrix of the static patch is represented in the CFT dual by a thermal state with imaginary temperature.

All of this can be viewed as  a space-to-time-rotated version of the presentation \cite{Maldacena:2001kr,Maldacena:2013xja} of AdS-Schwarzschild as a thermofield double and the Einstein-Rosen bridge emergent from their entanglement, much as dS/CFT itself resembles a space-to-time-rotated version of AdS/CFT. Similar observations to those of this paper, in a variety of contexts, have appeared in for example~\cite{Hawking:2000da,Narayan:2019pjl,Narayan:2017xca,Diaz:2019khq, Arias:2019lzk}. 

In this paper, we restrict for specificity to four-dimensional de Sitter space. Moreover, following \cite{Jafferis:2013qia}, we restrict the scalar mass to the limited range in which the $SO(4,1)$ conformal weights are real but non-integral. We expect these restrictions are non-essential. In Section~\ref{Sec:S2}, largely reviewing \cite{Jafferis:2013qia}, we describe a bulk  basis of (anti-)quasinormal scalar modes (QNMs), divided into two sets  which vanish in either the northern or southern static patch.  In Section~\ref{Sec:S3} we define a conjugate pair of bulk QNM quantum vacua (related by time reversal) as zero eigenstates of non-commuting operators. Two constructions of each dS $\alpha$-vacuum, including the Euclidean one, are given as squeezed-state excitations of these conjugate QNM vacua.  In Section~\ref{Sec:S4} we show that tracing the Euclidean vacuum over southern modes gives the expected northern density matrix $\rho_N$ proportional to $e^{-2\pi \textsf{H}_{\text{static}}}$, where $\textsf{H}_{\text{static}}$ is the static patch Hamiltonian. However, since the QNMs have imaginary frequencies, $\rho_N$ is not hermitian in the QNM basis with respect to the standard inner product compatible with the Klein-Gordon norm.
In Section~\ref{Sec:S5} we describe the dual boundary interpretation of these constructions. The tower of bulk QNMs in each patch are identified with a highest (or lowest) weight state and its descendants (or ascendants).  The boundary dilation operator `$\textsf{L}_0$' is identified with $-i$ times $\textsf{H}_{\text{static}}$\,, and the 4D bulk Euclidean vacuum is dual to a 2D boundary thermofield double state at temperature $\frac{1}{2\pi i}$. A simple formula is given which identifies the 4D scalar field operator at any bulk point with a linear combination of 2D operators which act on the thermofield double.  If the scalar field operator is at a point in the static patch, then it acts only on a single copy of the CFT$_3$.  A thermofield double state of two CFT$_3$ states is needed to describe global dS.  We conclude in Section~\ref{Sec:S6} with a brief discussion including a connection to CFT$_3$ disorder-averaged partition functions.

\section{Quasinormal modes}
\label{Sec:S2}

In this section we describe the mode expansion of a free massive scalar in terms of QNMs and AQNMs.  We will work in the dS$_4$ global coordinates $x=(t,\psi,\theta,\phi)$
with line element
\begin{equation}
\label{E:globalcoords1}
  \frac{ds^2}{\ell^2}
  =-dt^2+\cosh^2{t}\,[d\psi^2
    +\sin^2{\psi}\,(d\theta^2+\sin^2{\theta}d\phi^2)]\,,
\end{equation}
where $(\psi,\theta,\phi)$ are coordinates on the global $S^3$
slices, with the `north pole' at $\psi=\pi$ and `south pole' at $\psi=0$.  We associate the northern patch with a physical observer.  A massive scalar with wave equation
\begin{equation}
  (\square-m^2)\Phi=0
\end{equation}
has solutions in highest and lowest weight representations of the dS$_4$ $SO(4,1)$ isometry group.  
Highest (lowest) weight conformal primary solutions are given by \be \label{pmp}\Phi^{\pm \text{QN}}={ie^{-i \pi h_\pm} \over [\sinh t-\cosh t \cos\psi-i\epsilon ]^{h_\pm}}\,, ~~\Phi^{\pm \text{AQN}}={ie^{-i \pi h_\pm}  \over [\sinh t + \cosh t \cos\psi-i\epsilon ]^{h_\pm}}\,,\ee
where the conformal weight is 
\be h_\pm={3\over 2}\pm \sqrt{{9 \over 4}-m^2\ell^2}\,.\ee
We consider here  the range of $m$ in which $h_\pm$ are real,\footnote{We exclude the values for which $h_\pm$ are integral, where our construction becomes singular, although we do expect a suitably modified analysis to pertain both to this case and general complex weights.} and choose the phase so that denominators are real towards generic future points, but acquire phases $e^{ i \pi h_\pm}$ when crossing branch cuts towards the past. The overall phase convention in   \eqref{pmp} has been chosen so that
\be \Phi^{\pm \text{QN}}(\textsf{A}x)=- e^{-i \pi h_\pm}\Phi^{\pm \text{QN}*}(x)\,,\ee
where $\textsf{A}x \sim (-t, \pi-\psi,\pi-\theta,\phi-\pi) $ is antipodal to $x$. 

These solutions diagonalize the Lie action of the $SO(4,1)$ Killing vector 
\be \textsf{L}_0 =-\cos{\psi}\pd_t+\tanh{t}\sin{\psi}\pd_\psi  \ee
which is future timelike directed everywhere in the (northern) static patch.  One finds 
\be \textsf{L}_0 \Phi^{\pm \text{QN}}= - h_\pm \Phi^{\pm \text{QN}},  ~~~~ \textsf{L}_0 \Phi^{\pm \text{AQN}}= h_\pm \Phi^{\pm \text{AQN}},\ee
identifying $\Phi^{\pm \text{QN}}$ as QNMs and $\Phi^{\pm \text{AQN}}$ as AQNMs for the northern static patch.\footnote{Of course $\Phi^{\pm \text{QN}}$ are AQNMs for the southern static patch and vice versa.}$^{,}$\footnote{We note that our conventions differ from~\cite{Jafferis:2013qia} where $\Phi^{\pm \text{QN}}_{\text{us}} = \Phi^{\pm \text{AQN}}_{\text{them}}$ and $\Phi^{\pm \text{AQN}}_{\text{us}} = \Phi^{\pm \text{QN}}_{\text{them}}$ since we define the QNMs and AQNMs in relation to the northern static patch.}  In terms of northern static patch time $T = -\text{arctanh}(\sec \psi \, \tanh t )$, one has simply within the northern patch
\be \textsf{H}_{\text{static}}= i\,\textsf{L}_0=i\pd_T\,.\ee

In addition to $\textsf{L}_0$  there are three rotational  $\textsf{J}_k$ and 6 real boost $\textsf{M}_{\pm k}$ Killing vectors; see Appendix~\ref{App:Kvecs}. 
The highest and lowest weight conditions are
\be \textsf{M}_{-k}\Phi^{\pm \text{QN}}=\textsf{M}_{+k}\Phi^{\pm \text{AQN}}=0\,.\ee
The remaining $\textsf{M}_{\pm k}$ create descendants and ascendents which are also QNM or AQNMs. 
Taking all of these together we get the following eight towers of modes
\be \Phi_B^{\pm \text{QN}}, ~~\Phi_B^{\pm \text{QN}*},~~\Phi_{B}^{\pm \text{AQN}},~~\Phi_{B}^{\pm \text{AQN}*}, \ee
where $B$ is a multi-index indicating the action of powers of $\textsf{M}_{\pm k}$. 

It was shown in \cite{Jafferis:2013qia} that these eight towers are actually an overcomplete basis of solutions  of the wave equation: only half of them are needed. In the rest of this  paper we use only the  + modes $\Phi_B^{+\text{QN}}, ~\Phi_B^{+\text{AQN}}$ and their complex conjugates.\footnote{The $+$ and $-$  modes are related by a shadow transformation and at least in some cases form dual representations of the boundary CFT$_3$ \cite{Ng:2012xp}. In \cite{Jafferis:2013qia} a different collection of four towers were shown to provide a complete basis;  similar methods demonstrate completeness of the $+$ modes used here. }

\begin{figure}[t!]
\centering
\includegraphics[scale = .5]{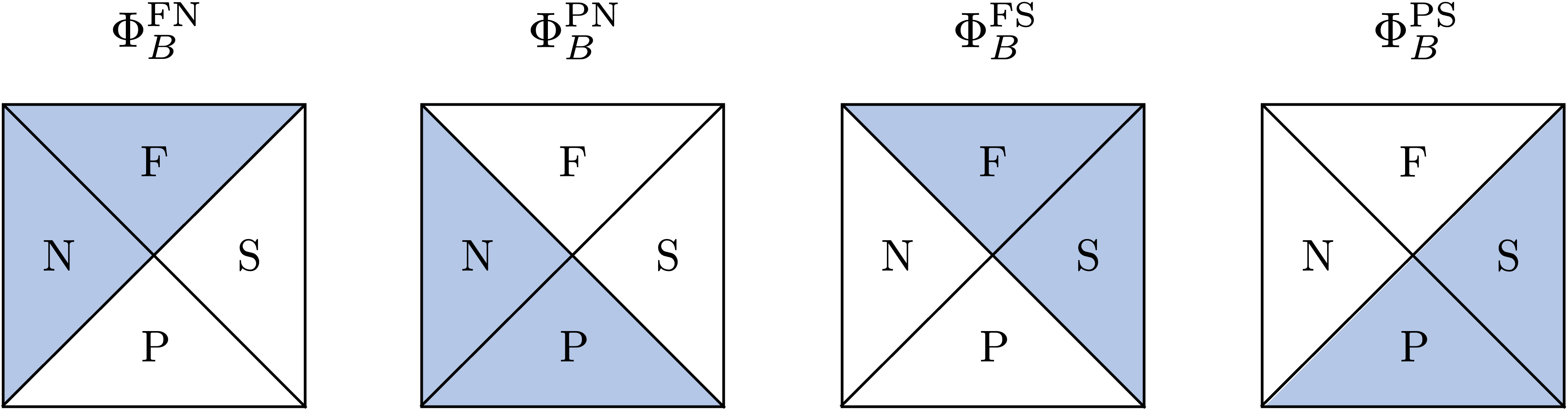}
\vspace{.4cm}
\caption{A depiction of the regions of support of the real, complete set of modes $\Phi_{B}^{\text{FN}}$,  $\Phi_{B}^{\text{PN}}$,  $\Phi_{B}^{\text{FS}}$,  $\Phi_{B}^{\text{PS}}$.  Here $\text{N}$ and $\text{S}$ denote the northern and southern static patches of global dS$_4$, and $\text{F}$ and $\text{P}$ denote the future and past Milne regions. \label{fig:fourmodes1}}
\end{figure}

We wish to construct northern and southern modes which have  no support in the southern and northern static patch respectively.
Northern modes are, dropping the + to avoid index clutter,  
\be  \label{oa} \Phi_{B}^{\text{FN}} =\Phi_{B}^{+\text{QN}} +\Phi_{B}^{+\text{QN}*},~~~~ \Phi_B^{\text{PN}} = -\left(e^{i \pi h_+}\Phi_B^{+\text{AQN}} +e^{-i \pi h_+}\Phi_B^{+\text{AQN}*}\right), \ee
while southern ones are 
\be \label{ob} \Phi_B^{\text{FS}} = \Phi_B^{+\text{AQN}} +\Phi_B^{+\text{AQN}*},~~~~ \Phi_{B}^{\text{PS}} =-\left(e^{i \pi h_+}\Phi_{B}^{+\text{QN}} +e^{-i \pi h_+}\Phi_{B}^{+\text{QN}*}\right). \ee
The inverses are \be \label{inv}\Phi^{+\text{QN}}_B={i \over 2\sin \pi h_+}(  e^{-i \pi h_+}\Phi_B^{\text{FN}} +\Phi_B^{\text{PS}})\,, ~~~~~  \Phi^{+\text{AQN}}_{B}={i \over 2\sin \pi h_+}(  e^{-i \pi h_+}\Phi_{B}^{\text{FS}} +\Phi_{B}^{\text{PN}})\,.\ee
In the following, we  employ the  complete set of manifestly real modes
\be \label{cmp} \Phi_B^{\text{FN}}, ~~\Phi_B^{\text{PN}},~~\Phi_{B}^{\text{FS}},~~\Phi_{B}^{\text{PS}}\,. \ee
These obey
\be \Phi_B^{\text{FS}}(x)=\Phi_B^{\text{PS}}(\textsf{T}x)=\Phi_B^{\text{FN}}(\textsf{P}x)=\Phi_B^{\text{PN}}(\textsf{A}x)\,, \ee
where $\textsf{T}x\sim (-t, \psi,\theta,\phi)$, $\textsf{P}x\sim (t, \pi-\psi,\pi-\theta,\phi-\pi)$ and $\textsf{A}=\textsf{PT}$. We note that $\Phi^{+\text{QN}}(\textsf{A}x)= - e^{-i\pi h_+}\Phi^{+\text{QN}*}(x)$. The region of support of these four types of modes is illustrated in Figure~\ref{fig:fourmodes1}.

Defining the Klein-Gordon inner product on global $S^3$
slices,
\begin{equation}
  \av{\Phi_1,\Phi_2}_{\text{KG}}
  := i\int_{S^3}\!d^3\Sigma^\mu\
  \Phi^{*}_{1}\overleftrightarrow{\pd_\mu}\Phi_{2}\,,
\end{equation}
we have the non-zero inner products
\be  \av{\Phi_A^{\text{FS}},  \Phi_B^{\text{PS}}}_{\text{KG}}=i\, N_{AB}=\av{\Phi_A^{\text{FN}},  \Phi_B^{\text{PN}}}_{\text{KG}}.\ee
Although we shall not need the explicit form of the  matrix  $N_{AB}$, it  can be computed using $SO(4,1)$ group theory (see~\cite{Jafferis:2013qia} and Appendix~\ref{App:Kvecs}) from that of the primaries (denoted $0$)
\be  \av{\Phi_0^{\text{FS}},  \Phi_0^{\text{PS}}}_{\text{KG}}= i\,8 \pi^{5/2} \frac{\tan(\pi h_+)}{\cos(\pi h_+)\,\Gamma(h_+)} \frac{\sin^2(\pi \mu)}{\Gamma(- \mu)}=\av{\Phi_0^{\text{FN}},  \Phi_0^{\text{PN}}}_{\text{KG}}\,,\ee
where here $\mu := \sqrt{\frac{9}{4} - m^2 \ell^2}$\,.  Note that we are using a different normalization than in~\cite{Jafferis:2013qia}.

\section{Vacua}
\label{Sec:S3}

In this Section we describe the construction of the Euclidean and other dS$_4$-invariant vacua.   Our starting point is the quasinormal mode  decomposition of the field operator
\be \label{rhj} \hat \Phi(x)= - i\,N^{AB}[\Phi^{\text{PN}}_A(x)\hat\Phi_B^{\text{FN}}-\Phi^{ \text{FN}}_A(x)\hat\Phi_B^{\text{PN}} + \Phi^{ \text{PS}}_A(x)\hat\Phi_B^{\text{FS}}-\Phi^{ \text{FS}}_A(x)\hat\Phi_B^{\text{PS}}]\ee
which, as we shall see, can be interpreted as a bulk reconstruction formula. 
In the above equation $N^{AB}N_{BC}=\delta^A_{C}$\,, the hat indicates an operator rather than a wavefunction, and 
\be \hat\Phi_B^{\text{FN}}= \big\langle  \hat \Phi, \Phi_B^{\text{FN}} \big\rangle_{\text{KG}}\,.\ee
The identity \eqref{rhj} can be shown by taking the Klein-Gordon inner product of both sides with the complete basis 
\eqref{cmp}.  The nontrivial commutators are 
\be\big[ \hat \Phi_A^{\text{FN}}, \hat \Phi_B^{\text{PN}}\big]=i\, N_{AB}=\big[ \hat \Phi_A^{\text{FS}}, \hat \Phi_B^{\text{PS}}\big]\,.\ee

Next we define a Rindler-like quasinormal vacuum state by
\be \label{sa} \hat\Phi_B^{\text{FN}}|0_{\text{F}}\rangle= \hat\Phi_B^{\text{FS}}|0_{\text{F}}\rangle=0\,.\ee
Since northern and southern operators commute we may decompose $|0_{\text{F}}\rangle=|0_{\text{FN}}\rangle|0_{\text{FS}}\rangle$.
Quantum states are then built as powers or functions of $\hat\Phi_B^{\text{PN}}$ and $\hat\Phi_B^{\text{PS}}$ acting on $| 0_{\text{F}}\rangle$. 
We may also define a second vacuum $|0_{\text{P}}\rangle$ obeying 
\be  \label{sb}\hat\Phi_B^{\text{PN}}|0_{\text{P}}\rangle= \hat\Phi_B^{\text{PS}}|0_{\text{P}}\rangle=0\,.\ee
Since they are annihilated by non-commuting hermitian operators,  $| 0_{\text{F}}\rangle $  and $| 0_{\text{P}}\rangle $ are non-normalizable states resembling zero position or zero momentum eigenstates. We may normalize them so that\footnote{It is useful to consider an analogy to the quantum mechanics of a particle in one spatial dimension: $|0_{\text{F}}\rangle$ is akin to the position eigenstate $|x = 0\rangle$, and $|0_{\text{P}}\rangle$ is akin to the momentum eigenstate $|p = 0\rangle$.  Then note that we may normalize the states so that $\langle p = 0 \,| x = 0 \rangle = 1$.}
\be \langle 0_{\text{P}}| 0_{\text{F}}\rangle =1\,.\ee
Two point functions may then be defined as 
\be\langle0_{\text{P}}| \hat \Phi_A^{\text{FN}} \hat \Phi_B^{\text{PN}}|0_{\text{F}}\rangle =i\,N_{AB}  =\langle0_{\text{P}}| \hat \Phi_A^{\text{FS}}\hat \Phi_B^{\text{PS}} |0_{\text{F}}\rangle\,. \ee
Since $\langle0_{\text{P}}| \hat \Phi(x) \hat \Phi(y)|0_{\text{F}}\rangle$ vanishes if  $x$ and $y$ are in opposite  patches,  there are no north-south correlations and  this does  not define  global quantum de Sitter correlators. Rather we have two decoupled theories, one for each patch.

Smooth global  dS$_4$-invariant   $\alpha$-vacua can nevertheless be constructed as squeezed-state-type excitations of either $| 0_{\text{F}}\rangle $  or $| 0_{\text{P}}\rangle $. They are 
defined by the condition
\begin{equation}\label{oc}
\left(\hat{\Phi}_B^{+\text{QN}\,\dagger} - e^{\alpha^*} \hat{\Phi}_B^{+\text{QN}}\right) |\alpha\rangle = \left(\hat{\Phi}_B^{+\text{AQN}\,\dagger} - e^{\alpha^*} \hat{\Phi}_B^{+\text{AQN}}\right) |\alpha\rangle = 0\,.
\end{equation}
It is readily shown that
\bea
|\alpha\rangle &\propto& e^{i \, \frac{\cosh(\alpha^*/2)}{\cosh(-\alpha^*/2 + i \pi h_+)} \,N^{AB} \hat{\Phi}_A^{\text{PS}} \hat{\Phi}_B^{\text{PN}}} |0_{\text{F}}\rangle\cr &\propto& e^{-i \, \frac{\cosh(-\alpha^*/2 + i \pi h_+)}{\cosh(\alpha^*/2)} \,N^{AB} \hat{\Phi}_A^{\text{FS}} \hat{\Phi}_B^{\text{FN}}} |0_{\text{P}}\rangle\,.
\eea
Note that the two representations of $|\alpha\rangle$ above need only be proportional to each other since~\eqref{oc} defines $|\alpha\rangle$ only up to a multiplicative prefactor.  The Euclidean vacuum corresponds to $\alpha^* \to - \infty$, in which case we find
\bea \label{sey}
|0_E\rangle&\propto &e^{i \, e^{-i \pi h_+} N^{AB} \hat{\Phi}_A^{\text{PS}} \hat{\Phi}_B^{\text{PN}}} |0_{\text{F}}\rangle\,\cr
&\propto&e^{-i \, e^{i \pi h_+} N^{AB} \hat{\Phi}_A^{\text{FS}} \hat{\Phi}_B^{\text{FN}}} |0_{\text{P}}\rangle\,.
\eea

It is also interesting to consider the singular limits $e^{\alpha^*}=-1$ and $e^{\alpha^*}=-e^{i2\pi h_+}$ which, comparing \eqref{oa}, \eqref{ob} and \eqref{oc}, formally correspond to \eqref{sa} and \eqref{sb}. In both cases the squeezed states \eqref{oc} diverge. Moreover from the general formula
\be G_\alpha(x,y)={ 1 \over  {1-e^{\alpha+\alpha^*}}}\big[G_E(x,y)+ e^{\alpha+\alpha^*}G_E(\textsf{A}x,\textsf{A}y)+e^{\alpha}G_E(\textsf{A}x,y)+ e^{\alpha^*}G_E(x,\textsf{A}y)\big]\,, \ee we see that the Green's functions also diverge. Hence this procedure does not produce a good global dS$_4$ vacuum from  the conditions \eqref{oa}, \eqref{ob}. Instead those conditions  describe two decoupled theories, one for each of the static patches.

\section{Northern density matrix}
\label{Sec:S4}

We wish to compute the density matrix obtained from a southern trace. Decomposing \bea |0_{\text{F}} \rangle&=&|0_{\text{FN}}\rangle|0_{\text{FS}}\rangle \\ |0_{\text{P}} \rangle&=& |0_{\text{PN}}\rangle|0_{\text{PS}}\rangle \nonumber \eea as the product of northern and southern vacua, \eqref{sey}  are analogs of the expression for the Minkowski vacuum as a Rindler thermofield double (in a basis with imaginary Rindler energies).  Let us now make an (unnormalized)  northern density matrix by tracing the Euclidean vacuum over the southern modes.  Letting  $\rho = |0_E\rangle \langle 0_E|$, which can be written as
\begin{align}
|0_E\rangle \langle 0_E| &\propto \sum_{m,n=0}^\infty \frac{(i \, e^{-i \pi h_+})^{m+n}}{m! \, n!} \,N^{A_1 C_1} \cdots N^{A_m C_m} \,N^{B_1 D_1} \cdots N^{B_n D_n}  \\
& \qquad \qquad \,\,\, \times   \hat{\Phi}_{A_1}^{\text{PN}} \cdots \hat{\Phi}_{A_m}^{\text{PN}} |0_{\text{FN}}\rangle \langle 0_{\text{PN}}| \hat{\Phi}_{B_1}^{\text{FN}} \cdots \hat{\Phi}_{B_n}^{\text{FN}} \otimes \hat{\Phi}_{C_1}^{\text{PS}} \cdots \hat{\Phi}_{C_m}^{\text{PS}} |0_{\text{FS}}\rangle \langle 0_{\text{PS}}| \hat{\Phi}_{D_1}^{\text{FS}} \cdots \hat{\Phi}_{D_n}^{\text{FS}}\,, \nonumber
\end{align}
we find that
\begin{align}
\label{E:rhoS1}
\rho_N := \text{tr}_S(\rho) \propto \sum_{n=0}^\infty \frac{(-i\,e^{-2  \pi i h_+})^n}{n!}\, N^{A_1 B_1} \cdots N^{A_n B_n} \,\hat{\Phi}_{A_1}^{\text{PN}} \cdots \hat{\Phi}_{A_n}^{\text{PN}} |0_{\text{FN}}\rangle \langle 0_{\text{PN}}| \hat{\Phi}_{B_n}^{\text{FN}} \cdots \hat{\Phi}_{B_1}^{\text{FN}}\,.
\end{align}
Since the right-hand side is unnormalized and proportional to $\rho_N$, let us denote it as $\widetilde{\rho}_N$.  To determine which operator this is, we observe that
\begin{equation}
\widetilde{\rho}_N \, \hat{\Phi}_{A_1}^{\text{PN}} \cdots \hat{\Phi}_{A_n}^{\text{PN}} |0_{\text{FN}}\rangle = e^{-2\pi i h_+ n} \hat{\Phi}_{A_1}^{\text{PN}} \cdots \hat{\Phi}_{A_n}^{\text{PN}} |0_{\text{FN}}\rangle\,,
\end{equation}
demonstrating that $\widetilde{\rho}_N = e^{-2\pi i \,\textsf{L}_0}$.  Letting $\textsf{H}_{\text{static}} =  i\,\textsf{L}_0 = i \,\partial_T$ where $T$ is the (northern) static patch time, we have
\begin{equation}
\rho_N \propto  e^{- 2\pi \textsf{H}_{\text{static}}}.
\end{equation}
This is the expected result, since we have merely done the usual southern trace in a different mode expansion.  The above suggests that the microstate-counting dS$_4$ partition function is the CFT$_3$ partition function  at temperature ${1 \over 2\pi i}$.
 
For $h_+$ non-integer, $\rho_N$ is not hermitian with respect to the standard Hilbert space inner product; this is a direct consequence of using a QNM basis, but the proper interpretation of this fact is not clear.  Notice that if $h_+$ is an integer, corresponding to $m^2 \ell^2 = 0$ or $2$ (as would be relevant e.g.~for Vasiliev gravity), then $\rho_N$ becomes the maximally mixed state.  However we have assumed here that  $h_+$ is not  an integer in order to avoid various divergences, so this case would have to be revisited with more care.  In any case, the possibility of $\rho_N$ being a maximally mixed state is highly suggestive, and is compatible with recent discussions of the static patch (see e.g.~\cite{Chandrasekaran:2022cip}).

\section{Dual boundary description}
\label{Sec:S5}

In dS$_4$/CFT$_3$, for each bulk  scalar field of mass $m$ there is a shadow pair of primary operators in the dual  CFT$_3$ with weights $h_\pm$. Here we consider the presentation of the CFT$_3$ in which the operator $\co$ has weight $h_+$. A tower of 3D states are then constructed by acting with (powers of) the operator $\co$ and all of its descendants $\co_B$. This matches the bulk states created by acting with (powers of) $\hat\Phi_B^{\text{PN}} $ on $|0_{F}\rangle$. However, in the bulk picture we have also the action of $\hat\Phi_B^{\text{PS}}$ which creates an isomorphic tower. Hence a state in the bulk Hilbert space lies in the tensor product of two boundary Hilbert spaces. This is different from usual AdS/CFT in global AdS, for which the bulk and boundary Hilbert spaces are identified. However, it matches exactly with the description of eternal AdS black holes as thermofield double states in AdS/CFT.

To be more precise, we consider two CFTs with vacuum state $|0\rangle_3$\,, and write
\begin{equation}
|A_1,...,A_n\rangle_3 := \frac{1}{\sqrt{n!}}\,\mathcal{O}_{A_1} \cdots \mathcal{O}_{A_n} |0\rangle_3\,.
\end{equation}
Here the `$3$' subscript emphasizes that we are considering a CFT$_3$.  Then we make the following identifications:
\begin{align}
\label{E:holo1}
\text{First CFT}_3\,:&\,\qquad \frac{1}{\sqrt{n!}}\,\hat{\Phi}_{A_1}^{\text{PN}} \cdots \hat{\Phi}_{A_n}^{\text{PN}} |0_{\text{FN}}\rangle \quad \longleftrightarrow \quad |A_1,...,A_n\rangle_3 \\
& \,\qquad \frac{1}{\sqrt{n!}}\,\langle 0_{\text{PN}}|\hat{\Phi}_{B_n}^{\text{FN}} \cdots \hat{\Phi}_{B_1}^{\text{FN}}  \quad \longleftrightarrow \quad \,_3\langle B_1,...,B_n|  \nonumber \\ \nonumber \\
\label{E:holo2}
\text{Second CFT}_3\,:&\,\qquad \frac{1}{\sqrt{n!}}\,\hat{\Phi}_{C_1}^{\text{PS}} \cdots \hat{\Phi}_{C_n}^{\text{PS}} |0_{\text{FS}}\rangle \quad \,\,\longleftrightarrow \quad |C_1,...,C_n\rangle_3 \quad \\
& \,\qquad \frac{1}{\sqrt{n!}} \langle 0_{\text{PS}}|\hat{\Phi}_{D_n}^{\text{FS}} \cdots \hat{\Phi}_{D_1}^{\text{FS}}  \quad \,\,\longleftrightarrow \quad \,_3\langle D_1,...,D_n| \,, \nonumber
\end{align}
and compute CFT$_3$ inner products with the standard BPZ norm. 
We see that  BPZ conjugation from bras to kets in the CFT$_3$s interchanges the bulk past and future.  Now we can write the holographic dual to the Euclidean vacuum as
\begin{equation}
\label{E:TFD1}
|\text{TFD}\rangle_3 \propto \sum_{n=0}^\infty e^{-i \pi  h_+ n} (i N^{A_1 C_1}) \cdots (i N^{A_n C_n})|A_1,...,A_n\rangle_3 \otimes |C_1,...,C_n\rangle_3\,,
\end{equation}
which is a thermofield double state between two CFTs with 3D imaginary temperature $\frac{1}{2\pi i}$\,.  Similarly, the reduced density matrix $\rho_N$ in~\eqref{E:rhoS1} can be expressed as a density matrix in a single CFT, namely
\begin{equation}
\label{E:rhoS2}
\rho_N \propto \sum_{n=0}^\infty e^{-2  \pi i h_+ n} (-i N^{A_1 B_1})\cdots (-i N^{A_n B_n}) \,|A_1,...,A_n\rangle_3\,_3\langle B_1,...,B_n|\,.
\end{equation}

More broadly, we are finding that the bulk Hilbert space of global dS$_4$ is isomorphic to two copies of a CFT$_3$ Hilbert space, or in an equation:
\begin{equation}
\mathcal{H}_{\text{global dS}_4} \simeq \mathcal{H}_{\text{CFT}_3} \otimes \mathcal{H}_{\text{CFT}_3}\,.
\end{equation}
It is the pattern of entanglement in~\eqref{E:TFD1} between the CFTs which builds up the bulk.  We expect this to generalize beyond the scalar field setting we have explored here.  The structure resembles the emergence of an AdS black hole spacetime from two copies of the CFT dual to AdS: the two static patches correspond to the two black hole exteriors, and their compliment corresponds to the AdS black hole and white hole regions. 

We can use our holographic correspondence above to rewrite bulk correlation functions in terms of CFT correlators.  For this purpose is convenient to introduce the notation
\begin{align}
\label{E:notation1}
|A_1,...,A_n\,; C_1,...,C_m\rangle_3 &:= (\mathcal{O}_{A_1} \cdots \mathcal{O}_{A_n}) \otimes (\mathcal{O}_{C_1} \cdots \mathcal{O}_{C_m}) |\text{TFD}\rangle_3 \\
\label{E:notation2}
|A_1,...,A_n\,; \varnothing \rangle_3 &:= (\mathcal{O}_{A_1} \cdots \mathcal{O}_{A_n}) \otimes \mathds{1} |\text{TFD}\rangle_3 \\
\label{E:notation3}
|\varnothing\,; C_1,...,C_m\rangle_3 &:= \mathds{1} \otimes (\mathcal{O}_{C_1} \cdots \mathcal{O}_{C_m}) |\text{TFD}\rangle_3 \\
\label{E:notation4}
|\varnothing\,; \varnothing \rangle_3 &:= |\text{TFD}\rangle_3\,.
\end{align}
As an example, consider the bulk 2-point function $\langle 0_E| \hat{\Phi}(x) \hat{\Phi}(y) |0_E\rangle$.  We can write it as
\newpage
\begin{align}
&\langle 0_E| \hat{\Phi}(x) \hat{\Phi}(y) |0_E\rangle \nonumber \\
& \qquad = \langle 0_E|  (-i\,N^{A_1 B_1})[\Phi^{ \text{PS}}_{A_1}(x)\hat\Phi_{B_1}^{\text{FS}}-\Phi^{ \text{FS}}_{A_1}(x)\hat\Phi_{B_1}^{\text{PS}}+\Phi^{\text{PN}}_{A_1}(x)\hat\Phi_{B_1}^{\text{FN}}-\Phi^{ \text{FN}}_{A_1}(x)\hat\Phi_{B_1}^{\text{PN}}] \nonumber \\
&\qquad \qquad \quad \quad \times (-i\,N^{A_2 B_2})[\Phi^{ \text{PS}}_{A_2}(y)\hat\Phi_{B_2}^{\text{FS}}-\Phi^{ \text{FS}}_{A_2}(y)\hat\Phi_{B_2}^{\text{PS}}+\Phi^{\text{PN}}_{A_2}(y)\hat\Phi_{B_2}^{\text{FN}}-\Phi^{ \text{FN}}_{A_2}(y)\hat\Phi_{B_2}^{\text{PN}}] |0_E\rangle \\ \nonumber \\ 
\label{E:bigsum1}
& \qquad = - N^{A_1 B_1} N^{A_2 B_2}\big( \Phi_{A_1}^{\text{PS}}(x) \Phi_{A_2}^{\text{PS}}(y) \langle 0_E| \hat{\Phi}_{B_1}^{\text{FS}} \hat{\Phi}_{B_2}^{\text{FS}}|0_E\rangle - \Phi_{A_1}^{\text{PS}}(x) \Phi_{A_2}^{\text{FS}}(y) \langle 0_E| \hat{\Phi}_{B_1}^{\text{FS}} \hat{\Phi}_{B_2}^{\text{PS}}|0_E\rangle \nonumber \\
& \qquad \qquad \qquad \qquad \qquad + \Phi_{A_1}^{\text{PS}}(x) \Phi_{A_2}^{\text{PN}}(y) \langle 0_E| \hat{\Phi}_{B_1}^{\text{FS}} \hat{\Phi}_{B_2}^{\text{FN}}|0_E\rangle - \Phi_{A_1}^{\text{PS}}(x) \Phi_{A_2}^{\text{FN}}(y) \langle 0_E| \hat{\Phi}_{B_1}^{\text{FS}} \hat{\Phi}_{B_2}^{\text{PN}}|0_E\rangle \nonumber \\
& \qquad \qquad \qquad \qquad \qquad -  \Phi_{A_1}^{\text{FS}}(x) \Phi_{A_2}^{\text{PS}}(y) \langle 0_E| \hat{\Phi}_{B_2}^{\text{FS}} \hat{\Phi}_{B_1}^{\text{PS}}|0_E\rangle + \Phi_{A_1}^{\text{FS}}(x) \Phi_{A_2}^{\text{FS}}(y) \langle 0_E| \hat{\Phi}_{B_1}^{\text{PS}} \hat{\Phi}_{B_2}^{\text{PS}}|0_E\rangle  \nonumber \\
& \qquad \qquad \qquad \qquad \qquad - \Phi_{A_1}^{\text{FS}}(x) \Phi_{A_2}^{\text{PN}}(y) \langle 0_E| \hat{\Phi}_{B_2}^{\text{FN}} \hat{\Phi}_{B_1}^{\text{PS}}|0_E\rangle + \Phi_{A_1}^{\text{FS}}(x) \Phi_{A_2}^{\text{FN}}(y) \langle 0_E| \hat{\Phi}_{B_1}^{\text{PS}} \hat{\Phi}_{B_2}^{\text{PN}}|0_E\rangle \nonumber \\
& \qquad \qquad \qquad \qquad \qquad + \Phi_{A_1}^{\text{PN}}(x) \Phi_{A_2}^{\text{PS}}(y) \langle 0_E| \hat{\Phi}_{B_1}^{\text{FN}} \hat{\Phi}_{B_2}^{\text{FS}}|0_E\rangle - \Phi_{A_1}^{\text{PN}}(x) \Phi_{A_2}^{\text{FS}}(y) \langle 0_E| \hat{\Phi}_{B_1}^{\text{FN}} \hat{\Phi}_{B_2}^{\text{PS}}|0_E\rangle  \nonumber \\
& \qquad \qquad \qquad \qquad \qquad + \Phi_{A_1}^{\text{PN}}(x) \Phi_{A_2}^{\text{PN}}(y) \langle 0_E| \hat{\Phi}_{B_1}^{\text{FN}} \hat{\Phi}_{B_2}^{\text{FN}}|0_E\rangle - \Phi_{A_1}^{\text{PN}}(x) \Phi_{A_2}^{\text{FN}}(y) \langle 0_E| \hat{\Phi}_{B_1}^{\text{FN}} \hat{\Phi}_{B_2}^{\text{PN}}|0_E\rangle  \nonumber \\
& \qquad \qquad \qquad \qquad \qquad - \Phi_{A_1}^{\text{FN}}(x) \Phi_{A_2}^{\text{PS}}(y) \langle 0_E| \hat{\Phi}_{B_2}^{\text{FS}} \hat{\Phi}_{B_1}^{\text{PN}}|0_E\rangle + \Phi_{A_1}^{\text{FN}}(x) \Phi_{A_2}^{\text{FS}}(y) \langle 0_E| \hat{\Phi}_{B_1}^{\text{PN}} \hat{\Phi}_{B_2}^{\text{PS}}|0_E\rangle  \nonumber \\
& \qquad \qquad \qquad \qquad \qquad - \Phi_{A_1}^{\text{FN}}(x) \Phi_{A_2}^{\text{PN}}(y) \langle 0_E| \hat{\Phi}_{B_2}^{\text{FN}} \hat{\Phi}_{B_1}^{\text{PN}}|0_E\rangle + \Phi_{A_1}^{\text{FN}}(x) \Phi_{A_2}^{\text{FN}}(y) \langle 0_E| \hat{\Phi}_{B_1}^{\text{PN}} \hat{\Phi}_{B_2}^{\text{PN}}|0_E\rangle \big) \nonumber \\
& \qquad \quad  \, + i \, N^{A_1 A_2} \left(\Phi_{A_1}^{\text{FS}}(x) \Phi_{A_2}^{\text{PS}}(y) + \Phi_{A_1}^{\text{FN}}(x) \Phi_{A_2}^{\text{PN}}(y)  \right) \langle 0_E| 0_E\rangle \\ \nonumber \\
\label{E:bigsum2}
& \qquad = - N^{A_1 B_1} N^{A_2 B_2}\big( \Phi_{A_1}^{\text{PS}}(x) \Phi_{A_2}^{\text{PS}}(y) \,_3\langle B_1, B_2\,; \varnothing | \varnothing\,;\varnothing\rangle_3 - \Phi_{A_1}^{\text{PS}}(x) \Phi_{A_2}^{\text{FS}}(y) \,_3\langle B_1 \,; \varnothing | B_2\,;\varnothing\rangle_3  \nonumber \\
& \qquad \qquad \qquad \qquad \qquad + \Phi_{A_1}^{\text{PS}}(x) \Phi_{A_2}^{\text{PN}}(y) \,_3\langle B_1\,; B_2\,|\varnothing\,;\varnothing\rangle_3 - \Phi_{A_1}^{\text{PS}}(x) \Phi_{A_2}^{\text{FN}}(y) \,_3\langle B_1\,;\varnothing | \varnothing\,;\,B_2\rangle_3 \nonumber \\
& \qquad \qquad \qquad \qquad \qquad -  \Phi_{A_1}^{\text{FS}}(x) \Phi_{A_2}^{\text{PS}}(y) \,_3\langle B_2\,;\varnothing | B_1\,;\varnothing \rangle_3 + \Phi_{A_1}^{\text{FS}}(x) \Phi_{A_2}^{\text{FS}}(y) \,_3\langle \varnothing\,;\varnothing | B_1, B_2\,;\varnothing\rangle_3  \nonumber \\
& \qquad \qquad \qquad \qquad \qquad - \Phi_{A_1}^{\text{FS}}(x) \Phi_{A_2}^{\text{PN}}(y) \,_3\langle \varnothing\,; B_2 | B_1\,;\varnothing\rangle_3  + \Phi_{A_1}^{\text{FS}}(x) \Phi_{A_2}^{\text{FN}}(y) \,_3\langle \varnothing\,;\varnothing | B_1\,; B_2 \rangle_3  \nonumber \\
& \qquad \qquad \qquad \qquad \qquad + \Phi_{A_1}^{\text{PN}}(x) \Phi_{A_2}^{\text{PS}}(y) \,_3\langle B_2\,; B_1 | \varnothing\,;\varnothing\rangle_3  - \Phi_{A_1}^{\text{PN}}(x) \Phi_{A_2}^{\text{FS}}(y) \,_3\langle \varnothing\,; B_1 | B_2\,;\varnothing\rangle_3   \nonumber \\
& \qquad \qquad \qquad \qquad \qquad + \Phi_{A_1}^{\text{PN}}(x) \Phi_{A_2}^{\text{PN}}(y) \,_3\langle \varnothing\,; B_1, B_2 | \varnothing\,;\varnothing\rangle_3  - \Phi_{A_1}^{\text{PN}}(x) \Phi_{A_2}^{\text{FN}}(y) \,_3\langle \varnothing\,; B_1 | \varnothing\,;B_2\rangle_3  \nonumber \\
& \qquad \qquad \qquad \qquad \qquad - \Phi_{A_1}^{\text{FN}}(x) \Phi_{A_2}^{\text{PS}}(y) \,_3\langle B_2\,; \varnothing | \varnothing\,;B_1\rangle_3 + \Phi_{A_1}^{\text{FN}}(x) \Phi_{A_2}^{\text{FS}}(y) \,_3\langle \varnothing\,; \varnothing | B_2\,;B_1\rangle_3  \nonumber \\
& \qquad \qquad \qquad \qquad \qquad - \Phi_{A_1}^{\text{FN}}(x) \Phi_{A_2}^{\text{PN}}(y) \,_3\langle \varnothing\,; B_2 | \varnothing\,;B_1\rangle_3 + \Phi_{A_1}^{\text{FN}}(x) \Phi_{A_2}^{\text{FN}}(y) \,_3\langle \varnothing\,; \varnothing | \varnothing\,; B_1, B_2\rangle_3 \big) \nonumber \\
& \qquad \quad  \, + i \, N^{A_1 A_2} \left(\Phi_{A_1}^{\text{FS}}(x) \Phi_{A_2}^{\text{PS}}(y) + \Phi_{A_1}^{\text{FN}}(x) \Phi_{A_2}^{\text{PN}}(y)  \right) \,_3\langle \varnothing\,;\varnothing| \varnothing\,;\varnothing \rangle_3\,.
\end{align}
Equation~\eqref{E:bigsum1} expresses the 2-point function as a sum of correlators where we have used the commutation relations so that those correlators containing both `future' and `past' operators have the `future' operators on the left and the `past' operators on the right.  As such, we can use~\eqref{E:holo1},~\eqref{E:holo2} in tandem with~\eqref{E:notation1},~\eqref{E:notation2},~\eqref{E:notation3},~\eqref{E:notation4} to rewrite~\eqref{E:bigsum1} in terms of expectation values of tensor products of CFT$_3$ operators in the thermofield double state.  This gives us the expression in~\eqref{E:bigsum2}.  The above approach readily generalizes to higher-point correlation functions.

What is the bulk dual of a single copy of the CFT$_3$? The modes $\Phi_B^{\text{PN}}$ and their conjugates $\Phi_B^{\text{FN}} $ form a basis of functions in the northern static patch where the southern modes simply vanish.  Hence bulk particles in the northern static patch can be written as CFT Fock space states created by (products) of the $\co_A$.  This means that the Hilbert space of the single static patch is contained within the Hilbert space of a single CFT$_3$:
\begin{equation}
\mathcal{H}_{\text{static}} \subset \mathcal{H}_{\text{CFT}_3}\,.
\end{equation}
If we take the CFT$_3$ to be described by the density matrix $\rho_N$ in~\eqref{E:rhoS2} which has 3D temperature $\frac{1}{2\pi i}$\,, then for any point $x$ in the northern static patch the bulk reconstruction formula \eqref{rhj} reduces to 
\be \label{rhj2} \hat \Phi(x)= - i\,N^{AB}[\Phi^{ \text{PN}}_A(x)\hat\Phi_B^{\text{FN}}-\Phi^{ \text{FN}}_A(x)\hat\Phi_B^{\text{PN}}]\,.\ee 
Hence  a single copy of the boundary CFT$_3$ in a mixed state can reproduce the bulk static patch correlators in the Euclidean vacuum.

\section{Discussion}
\label{Sec:S6}

Although many aspects of the AdS/CFT dictionary are well-understood,  the dS/CFT dictionary is still largely under development.  Our work here emphasizes the importance of understanding dS/CFT for global de Sitter, for which there is evidently an interplay between multiple CFTs (see~\cite{Cotler:2019nbi, Cotler:2019dcj, Doi:2022iyj} for recent discussions).  In particular, we find that the entanglement between two CFTs gives rise to the bulk fields and geometry.

\begin{figure}[t]
\centering
\includegraphics[scale=.45]{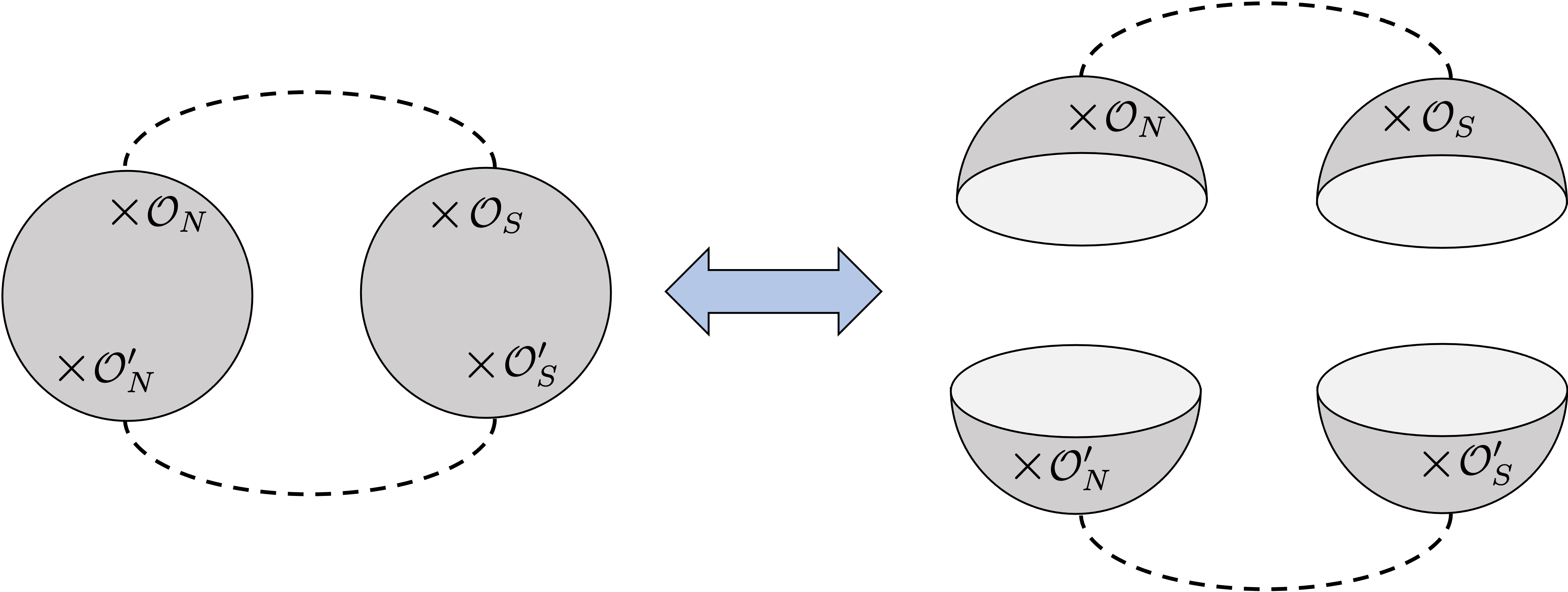}
\caption{The left-hand side depicts two $S^3$ partition functions (with operator insertions) of two CFTs, jointly disorder-averaged over some source to which they are coupled (dotted lines).  The right-hand side depicts what happens when we break the spheres along an $S^2$ equator, arriving at two states which are each entangled between two CFTs.  Here the dotted lines denote the entanglement between the CFTs. \label{fig:EEdisorder}}
\end{figure}

Here we mention  another perspective on dS/CFT for global de Sitter that is connected with our findings.  Recent work~\cite{Cotler:2019nbi, Cotler:2019dcj} has hypothesized that the dual to global de Sitter can be described by two CFTs which are coupled by a disorder average.  Schematically, let $Z_{S^3}^{\mathcal{O}, \mathcal{O}'}[J]$ be a CFT$_3$ partition function on $S^3$, where we have inserted some operators $\mathcal{O}$, $\mathcal{O}'$ on the sphere and coupled the CFT to some source $J$.  Then correlation functions in the dS$_4$ bulk are hypothesized to be equivalent to the 2-replica quantity
\begin{equation}
\label{E:ZZeq1}
\Big\langle Z_{S^3}^{\mathcal{O}_N, \mathcal{O}_N'}[J] \,Z_{S^3}^{\mathcal{O}_S, \mathcal{O}_S'}[J] \Big\rangle_{J}
\end{equation}
where $\langle \, \cdot \, \rangle_J$ denotes some disorder averaging over $J$'s.  Here we have suggestively labeled the operators in the first CFT with N subscripts (for `north') and labeled the operators in the second CFT with S subscripts (for `south').  Equation~\eqref{E:ZZeq1} is depicted on the left-hand side of Figure~\ref{fig:EEdisorder}, where the dotted lines denote the $J$ disorder average.  Considering the same Figure, if we cut each $S^3$ into two halves along an $S^2$, then we arrive at the picture on the right-hand side: we have two entangled states between the two CFTs, where now the dotted lines denote the entanglement between the CFTs.  Denoting the states by $\langle \Psi_{\mathcal{O}_N, \mathcal{O}_S}|$ and $| \Psi_{\mathcal{O}_{N}', \mathcal{O}_S'}\rangle$, we have
\begin{equation}
\Big\langle Z_{S^3}^{\mathcal{O}_N, \mathcal{O}_N'}[J] \,Z_{S^3}^{\mathcal{O}_S, \mathcal{O}_S'}[J] \Big\rangle_{J} = \langle \Psi_{\mathcal{O}_N, \mathcal{O}_S} | \Psi_{\mathcal{O}_N', \mathcal{O}_S'}\rangle\,.
\end{equation}
As such, the disorder-averaging approach may be related to our entanglement approach.\footnote{We note that in~\cite{Cotler:2019nbi, Cotler:2019dcj} the two disordered CFTs were respectively associated with the far past and far future of global de Sitter.  Here it appears more natural for the two disordered CFTs to be respectively associated with the northern and southern static patches.}

Our work has strongly leveraged the quantization of quasinormal modes.  The quasinormal mode basis has some unusual properties, for instance having imaginary eigenvalues with respect to the static patch Hamiltonian.  This leads to technical and conceptual questions regarding hermiticity, as manifested by our computation of $\rho_N$.  It would be valuable to better understand the hermiticity or non-hermiticity inherent in this quantization, and explore the interplay with fields for which $h_+$ in an integer.

Finally, our work suggests that quantum entanglement between two CFTs may be in part responsible for the emergence of time in global de Sitter.  This resonates with recent work~\cite{Leutheusser:2021frk, Witten:2021unn, Leutheusser:2022bgi} in the AdS setting.  In the dS/CFT setting, perhaps entanglement between the two CFTs can better elucidate the code subspace discussed in~\cite{Cotler:2022weg}.

\section*{Acknowledgements}

We thank Kristan Jensen and Noah Miller for valuable discussions. We also thank Anthonny Canazas for pointing out a typo in an earlier version of the paper.  JC is supported by a Junior Fellowship from the Harvard Society of Fellows.  AS is supported in part by DOE grant de-sc/0007870.

\appendix

\section{dS$_4$ Killing vectors}  
\label{App:Kvecs}

Using the dS$_4$ global coordinates in~\eqref{E:globalcoords1}, the 10 Killing vectors are\footnote{Our Killing vectors are related to those of~\cite{Jafferis:2013qia} by $t \to -t$.}
\begin{align}
\textsf{L}_0 =\,& -\cos \psi \, \partial_t + \tanh t \, \sin \psi \, \partial_\psi \\
\textsf{M}_{\pm 1} =\,& \pm \sin \psi \, \sin \theta \, \sin \phi \,\partial_t + (1 \pm \tanh t \, \cos \psi) \sin \theta \, \sin \phi \, \partial_\psi \\
& + (\cot \psi \pm \tanh t \, \csc \psi )(\cos \theta \, \sin \phi \, \partial_\theta + \csc\theta \, \cos\phi \, \partial_\phi) \nonumber  \\
\textsf{M}_{\pm 2} =\,& \pm \sin \psi \, \sin \theta \, \sin \phi \,\partial_t + (1 \pm \tanh t \, \cos \psi) \sin \theta \, \sin \phi \, \partial_\psi \\
& + (\cot \psi \pm \tanh t \, \csc \psi )(\cos \theta \, \cos \phi \, \partial_\theta - \csc\theta \, \sin\phi \, \partial_\phi) \nonumber  \\
\textsf{M}_{\pm 3} =\,& \pm \sin \psi \, \cos \theta \, \partial_t + (1 \pm \tanh t \, \cos \psi) \cos \theta \, \partial_\psi - (\cot \psi \pm \tanh t \, \csc \psi) \sin \theta \, \partial_\theta \\
\textsf{J}_1 =\,& \cos \phi \, \partial_\theta - \sin \phi \, \cot \theta \, \partial_\phi \\
\textsf{J}_2 =\,& -\sin \phi \, \partial_\theta - \cos \phi \, \cot \theta \, \partial_\phi \\
\textsf{J}_3 = \,& \partial_\phi
\end{align}
with the non-zero commutators
\begin{align}
[\textsf{L}_0\,,\textsf{M}_{\pm i}] &= \mp \textsf{M}_{\pm i}\,, \qquad \qquad  [\textsf{M}_{+i}\,,\textsf{M}_{-j}] = 2 \textsf{L}_0 \, \delta_{ij} + 2 \sum_{k=1}^3 \epsilon_{ijk}\,\textsf{J}_k \nonumber \\
[\textsf{J}_i\,,\textsf{J}_j] &= \sum_{k=1}^3 \epsilon_{ijk}\,\textsf{J}_k\,, \qquad \quad \,\,  [\textsf{J}_i\,, \textsf{M}_{\pm j}] = \sum_{k=1}^3 \epsilon_{ijk}\,\textsf{M}_{\pm k}\,.
\end{align}
This forms a representation of $SO(4,1)$.

\bibliography{ds.bib}

\providecommand{\href}[2]{#2}\begingroup\raggedright\begin{thebibliography}{10}

\bibitem{Ryu:2006bv}
S.~Ryu and T.~Takayanagi, ``{Holographic derivation of entanglement entropy
  from AdS/CFT},'' \href{http://dx.doi.org/10.1103/PhysRevLett.96.181602}{{\em
  Phys. Rev. Lett.} {\bfseries 96} (2006) 181602},
  \href{http://arxiv.org/abs/hep-th/0603001}{{\ttfamily arXiv:hep-th/0603001}}.

\bibitem{VanRaamsdonk:2010pw}
M.~Van~Raamsdonk, ``{Building up spacetime with quantum entanglement},''
  \href{http://dx.doi.org/10.1142/S0218271810018529}{{\em Gen. Rel. Grav.}
  {\bfseries 42} (2010) 2323--2329},
  \href{http://arxiv.org/abs/1005.3035}{{\ttfamily arXiv:1005.3035 [hep-th]}}.

\bibitem{Maldacena:2013xja}
J.~Maldacena and L.~Susskind, ``{Cool horizons for entangled black holes},''
  \href{http://dx.doi.org/10.1002/prop.201300020}{{\em Fortsch. Phys.}
  {\bfseries 61} (2013) 781--811},
  \href{http://arxiv.org/abs/1306.0533}{{\ttfamily arXiv:1306.0533 [hep-th]}}.

\bibitem{Almheiri:2014lwa}
A.~Almheiri, X.~Dong, and D.~Harlow, ``{Bulk Locality and Quantum Error
  Correction in AdS/CFT},''
  \href{http://dx.doi.org/10.1007/JHEP04(2015)163}{{\em JHEP} {\bfseries 04}
  (2015) 163}, \href{http://arxiv.org/abs/1411.7041}{{\ttfamily arXiv:1411.7041
  [hep-th]}}.

\bibitem{Garfinkle:1993xk}
D.~Garfinkle, S.~B. Giddings, and A.~Strominger, ``{Entropy in black hole pair
  production},'' \href{http://dx.doi.org/10.1103/PhysRevD.49.958}{{\em Phys.
  Rev. D} {\bfseries 49} (1994) 958--965},
  \href{http://arxiv.org/abs/gr-qc/9306023}{{\ttfamily arXiv:gr-qc/9306023}}.

\bibitem{Maldacena:2001kr}
J.~M. Maldacena, ``{Eternal black holes in anti-de Sitter},''
  \href{http://dx.doi.org/10.1088/1126-6708/2003/04/021}{{\em JHEP} {\bfseries
  04} (2003) 021}, \href{http://arxiv.org/abs/hep-th/0106112}{{\ttfamily
  arXiv:hep-th/0106112}}.

\bibitem{Strominger:2001pn}
A.~Strominger, ``{The dS / CFT correspondence},''
  \href{http://dx.doi.org/10.1088/1126-6708/2001/10/034}{{\em JHEP} {\bfseries
  10} (2001) 034}, \href{http://arxiv.org/abs/hep-th/0106113}{{\ttfamily
  arXiv:hep-th/0106113}}.

\bibitem{Ng:2012xp}
G.~S. Ng and A.~Strominger, ``{State/Operator Correspondence in Higher-Spin
  dS/CFT},'' \href{http://dx.doi.org/10.1088/0264-9381/30/10/104002}{{\em
  Class. Quant. Grav.} {\bfseries 30} (2013) 104002},
  \href{http://arxiv.org/abs/1204.1057}{{\ttfamily arXiv:1204.1057 [hep-th]}}.

\bibitem{Jafferis:2013qia}
D.~L. Jafferis, A.~Lupsasca, V.~Lysov, G.~S. Ng, and A.~Strominger,
  ``{Quasinormal quantization in de Sitter spacetime},''
  \href{http://dx.doi.org/10.1007/JHEP01(2015)004}{{\em JHEP} {\bfseries 01}
  (2015) 004}, \href{http://arxiv.org/abs/1305.5523}{{\ttfamily arXiv:1305.5523
  [hep-th]}}.

\bibitem{Hawking:2000da}
S.~Hawking, J.~M. Maldacena, and A.~Strominger, ``{de Sitter entropy, quantum
  entanglement and AdS / CFT},''
  \href{http://dx.doi.org/10.1088/1126-6708/2001/05/001}{{\em JHEP} {\bfseries
  05} (2001) 001}, \href{http://arxiv.org/abs/hep-th/0002145}{{\ttfamily
  arXiv:hep-th/0002145}}.

\bibitem{Narayan:2019pjl}
K.~Narayan, ``{de Sitter entropy as entanglement},''
  \href{http://dx.doi.org/10.1142/S021827181944019X}{{\em Int. J. Mod. Phys. D}
  {\bfseries 28} no.~14, (2019) 1944019},
  \href{http://arxiv.org/abs/1904.01223}{{\ttfamily arXiv:1904.01223
  [hep-th]}}.

\bibitem{Narayan:2017xca}
K.~Narayan, ``{On extremal surfaces and de Sitter entropy},''
  \href{http://dx.doi.org/10.1016/j.physletb.2018.02.010}{{\em Phys. Lett. B}
  {\bfseries 779} (2018) 214--222},
  \href{http://arxiv.org/abs/1711.01107}{{\ttfamily arXiv:1711.01107
  [hep-th]}}.

\bibitem{Diaz:2019khq}
F.~Diaz, ``{de Sitter Entanglement and Conformal Description of the
  Cosmological Horizon},'' Master's thesis, Andres Bello Natl. U., 2019.

\bibitem{Arias:2019lzk}
C.~Arias, F.~Diaz, and P.~Sundell, ``{Gibbons\textendash{}Hawking entropy as
  entanglement entropy},'' \href{http://dx.doi.org/10.1063/1.5130124}{{\em AIP
  Conf. Proc.} {\bfseries 2163} no.~1, (2019) 090002}.

\bibitem{Chandrasekaran:2022cip}
V.~Chandrasekaran, R.~Longo, G.~Penington, and E.~Witten, ``{An Algebra of
  Observables for de Sitter Space},''
  \href{http://arxiv.org/abs/2206.10780}{{\ttfamily arXiv:2206.10780
  [hep-th]}}.

\bibitem{Cotler:2019nbi}
J.~Cotler, K.~Jensen, and A.~Maloney, ``{Low-dimensional de Sitter quantum
  gravity},'' \href{http://dx.doi.org/10.1007/JHEP06(2020)048}{{\em JHEP}
  {\bfseries 06} (2020) 048}, \href{http://arxiv.org/abs/1905.03780}{{\ttfamily
  arXiv:1905.03780 [hep-th]}}.

\bibitem{Cotler:2019dcj}
J.~Cotler and K.~Jensen, ``{Emergent unitarity in de Sitter from matrix
  integrals},'' \href{http://dx.doi.org/10.1007/JHEP12(2021)089}{{\em JHEP}
  {\bfseries 12} (2021) 089}, \href{http://arxiv.org/abs/1911.12358}{{\ttfamily
  arXiv:1911.12358 [hep-th]}}.

\bibitem{Doi:2022iyj}
K.~Doi, J.~Harper, A.~Mollabashi, T.~Takayanagi, and Y.~Taki, ``{Pseudo Entropy
  in dS/CFT and Time-like Entanglement Entropy},''
  \href{http://arxiv.org/abs/2210.09457}{{\ttfamily arXiv:2210.09457
  [hep-th]}}.

\bibitem{Leutheusser:2021frk}
S.~Leutheusser and H.~Liu, ``{Emergent times in holographic duality},''
  \href{http://arxiv.org/abs/2112.12156}{{\ttfamily arXiv:2112.12156
  [hep-th]}}.

\bibitem{Witten:2021unn}
E.~Witten, ``{Gravity and the crossed product},''
  \href{http://dx.doi.org/10.1007/JHEP10(2022)008}{{\em JHEP} {\bfseries 10}
  (2022) 008}, \href{http://arxiv.org/abs/2112.12828}{{\ttfamily
  arXiv:2112.12828 [hep-th]}}.

\bibitem{Leutheusser:2022bgi}
S.~Leutheusser and H.~Liu, ``{Subalgebra-subregion duality: emergence of space
  and time in holography},'' \href{http://arxiv.org/abs/2212.13266}{{\ttfamily
  arXiv:2212.13266 [hep-th]}}.

\bibitem{Cotler:2022weg}
J.~Cotler and A.~Strominger, ``{The Universe as a Quantum Encoder},''
  \href{http://arxiv.org/abs/2201.11658}{{\ttfamily arXiv:2201.11658
  [hep-th]}}.

\end{thebibliography}\endgroup
\bibliographystyle{utphys}

\end{document}